\documentclass[smallextended,natbib]{svjour3_arXiv}  

\smartqed  
\usepackage{graphicx}
\usepackage{xcolor}
\usepackage[colorlinks=true,
			linkcolor = blue,
            urlcolor  = blue,
            citecolor = blue,
            anchorcolor = blue]{hyperref}
\usepackage{xspace}
\usepackage{amssymb,amsmath,amsfonts}
\usepackage{mathtools}
\usepackage{chemarrow,dsfont}
\usepackage{listings}

\newcommand{\Aboria}{\href{https://martinjrobins.github.io/Aboria/index.html}{Aboria}\xspace}

\newcommand{\ud}{\mathrm{d}}
\newcommand{\bfx}{{\bf x}}
\newcommand{\bfp}{{\bf p}}
\newcommand{\bfX}{{\bf X}}

\journalname{Bulletin of Mathematical Biology}
\begin{document}

\title{Particle-based simulations of reaction-diffusion processes with Aboria
\thanks{This work was supported by EPSRC (grant EP/I017909/1), St John's College Research Centre (grant 21138701) and the John Fell Fund (grant BLD10370).}
}
\subtitle{Application to a heterogeneous population of cells with chemotaxis and volume exclusion}


\author{Maria Bruna \and Philip~K. Maini \and Martin Robinson}


\institute{M. Bruna \and P.~K. Maini \at
 Mathematical Institute, University of Oxford, Radcliffe Observatory Quarter, Woodstock Road, Oxford, OX2 6GG, United Kingdom\\
              \email{bruna@maths.ox.ac.uk}, \email{maini@maths.ox.ac.uk}           
           \and
           M. Robinson \at
              Department of Computer Science, University of Oxford, Wolfson 
    Building, Parks Rd, Oxford OX1 3QD, United Kingdom\\
    \email{martin.robinson@cs.ox.ac.uk}
}


\maketitle

\begin{abstract} 
Mathematical models of transport and reactions in biological systems have been traditionally written in terms of partial differential equations (PDEs) that describe the time evolution of population-level variables.
In recent years, the use of stochastic particle-based  models, which keep track of the evolution of each organism in the system, has become widespread.
These models provide a lot more detail than the population-based PDE models, for example by explicitly modelling particle-particle interactions, but bring with them many computational challenges. 
In this paper we overview \Aboria, a powerful and flexible C\texttt{++} library for the implementation of numerical methods for particle-based models. 
We demonstrate the use of Aboria with a commonly used model in mathematical biology, namely cell chemotaxis. Cells interact with each other and diffuse, biased by extracellular chemicals, that can be altered by the cells themselves. We use a hybrid approach where particle-based models of cells are coupled with a PDE for the concentration of the extracellular chemical.
\end{abstract}

\section{Introduction}
\label{sec:intro}

Biology has recently experienced a revolution through the development of new,
quantitative, measurement techniques. As an example, super-resolution microscopy
allows biomolecules to be localised, counted and distinguished
\citep{Betzig:2006hx}. Traditional modelling approaches using partial
differential equations (PDEs) for population-level variables (concentrations,
densities) are unable to capture and explain some of the particle-level
(for example, molecule- or cell-level) features that we are now obtaining from experiments. This
is why a more detailed modelling approach, namely particle-based or agent-based
models that describe the behaviour of each organism, is now increasingly used by the
mathematical biology community. Computer simulations are ideal for studying the
dynamical mechanisms arising from the interplay of these particles, filling
in the details that cannot be resolved experimentally, and testing/generating biological hypothesis. These simulations
comprise algorithms for particle-based stochastic reaction-diffusion processes
that track individual particles and are computationally expensive. 

The computational requirements of particle-based models, such as the efficient
calculation of interactions between particles, are challenging to implement in a
way that scales well with the number of particles, uniform and non-uniform
particle distributions, different spatial dimensions and periodicity. The
majority of existing software is typically designed to suit a particular application, and therefore tends not to be sufficiently generic that can be used to implement these computationally challenging routines,
and so they are reimplemented again and again in each software package.

The primary class of software for particle-based methods is geared towards molecular dynamics simulations, such as the hugely popular 
\href{http://www.gromacs.org/}{GROMACS} \citep{abraham2015gromacs}, or
\href{http://openmm.org/}{OpenMM} \citep{eastman2012openmm} packages. These
generally include sophisticated neighbour searching and evaluation of long-range
forces, but do not include the possibility of reactions between particles, a
vital component of many biological models. On the other hand, the biochemical
simulator package \href{http://www.smoldyn.org/}{Smoldyn} \citep{Andrews:2004fs} implements both unimolecular and bimolecular reactions
using the Smoluchowski method, but has limited capabilities to implement interactions between particles and transport mechanisms other than unbiased Brownian motion.

\Aboria is a C\texttt{++} library for the numerical implementation of
particle-based models \citep{Robinson:2017vxa}. Aboria aims to provide a general
purpose library that allows the user complete control to define whatever
interactions they wish on a given particle set, while implementing efficiently
the difficult algorithmic aspects of particle-based methods, such as
neighbourhood searches and the calculation of long-range forces. It is agnostic
to any particular numerical method, for example it does not contain any
particular molecular dynamics or Smoluchowski dynamics algorithms, but instead
provides the user with computational tools that allow them to implement these
algorithms in a customised fashion. It is thus especially suitable for the
implementation of novel particle-based methods, or hybrid models that couple
particle-based models with continuum PDE models. In particular, Aboria gives the
user:
\begin{enumerate}
\item A container class to store a particle set with a position and unique id
for each particle, as well as any number of user-defined variables associated to
each particle with arbitrary types (for example, to store a velocity, a force,
or the particle size).

\item The ability to embed each particle set within an $n$-dimensional hypercube with arbitrary periodicity. The underlying data
structure can be a cell list, kd-tree or hyper oct-tree. The data structure can
be chosen to suit the particular application.

\item Flexible neighbourhood queries that return iterators, and can use any
integer $p$-norm distance measure (for example, the Euclidean distance is
given by $p = 2$, the Chebyshev distance is given by $p= \infty$). 

\item The ability to calculate long-range forces between $N$ particles
efficiently (that is, in $O(N)$) using the black-box fast multipole method
\citep{fong2009black}. This method can be used for any long-range force that is
well-approximated by a low-order polynomial for large distances between
particles (the computational cost of the model scales with the degree of the
polynomial). 

\item An easy to use embedded Domain Specific Language (eDSL) expressing a wide
variety of particle-particle interactions. For specialised cases which cannot be
easily expressed by this symbolic eDSL, Aboria also provides a low-level
interface based on the C\texttt{++} Standard Template Library
\citep{stroustrup2013cpp}. 

\end{enumerate}

The aim of this paper is to demonstrate how Aboria can be used to implement many
features of particle-based models common in mathematical biology. These include
heterogeneous particle populations, interactions between particles, volume
exclusion, biased transport, reactions, and proliferation. We showcase the many
features of the Aboria library through a well-known model in mathematical
biology, namely cell chemotaxis. In Sect.~\ref{sec:Model} we detail the
mathematical model for chemotaxis. Then in Sect.~\ref{sec:Implementation} we
show a step-by-step implementation of the model with Aboria. Finally, we show
how to output and analyse the data using Python in Sect.~\ref{sec:Results}. 

\section{Model for chemotaxis}
\label{sec:Model}

As our guiding example, we consider chemotactic cell migration with volume
exclusion. In particular, we study a population of $N$ cells divided into two
sub-populations (types $\alpha$ and $\beta$) and a diffusing attractive chemical
substance that is produced by cells of type $\alpha$ and consumed by cells of
type $\beta$. Cells move around in a two-dimensional domain $\Omega$ due to Brownian motion,
biased by gradients in the concentration of the chemical in the case of type
$\beta$ cells. Cells can also undergo reactions that change their type from
chemical producers to chemical consumers. Volume exclusion can be modelled by either point particles with a short-range repulsive interaction potential $u$ (so that cells are allowed to deform if in close contact) or hard-sphere particles that are not allowed to overlap. To make the example
as broad as possible, we take a hybrid modelling approach whereby the cells are
modelled discretely using particle-based models for Brownian motion, while the
chemical is represented by its concentration $c$ using a reaction-diffusion PDE.
Hybrid models of chemotaxis have been considered in
\cite{Guo:2008ec,Dallon:1997be,Franz:2013ds,Newman:2004cj,McLennan:2012jb}.

\begin{subequations}
\label{model}	

Let $\bfX_i(t)$ denote the position of the $i$th particle in $\Omega \subset \mathbb R^2$. Let
$S_\alpha(t)$ denote the set of cells of type $\alpha$ and $S_\beta(t)$
denote the set of cells of type $\beta$ at time $t$. For each particle, the
motion through space is described by a stochastic differential equation (SDE).
For $i \in S_\alpha(t)$, 
\begin{equation}
	\ud \bfX_i(t) = \sqrt{2 D_\alpha} \ud {\bf W}_i(t) - \sum_{j\ne i} \nabla_i u(\| \bfX_i(t) - \bfX_j(t) \|) \ud t,
	\label{eq:sde_alpha} 
\end{equation}
where $D_\alpha$ is the diffusion coefficient of cells of type $\alpha$, and
$\nabla_i$ denotes the gradient with respect to $\bfX_i$. For $i \in S_\beta(t)$
\begin{equation}
	\ud \bfX_i(t) = \sqrt{2D_\beta} \ud{ \bf W}_i(t) + \chi \nabla c(\bfX_i(t),t) \ud t - \sum_{j\ne i} \nabla_i u(\| \bfX_i(t) - \bfX_j(t) \|) \ud t.\label{eq:sde_beta} 
\end{equation}
Here $D_\beta$ is the diffusion coefficient of cells of type $\beta$, $\chi$
denotes the chemotactic sensitivity of the cells (taken to be
constant) and $c(\bfx,t)$ is the concentration of chemical at position $\bfx$ and
time $t$. The interaction potential $u$ may be a soft potential
incorporating effects such as size exclusion by cells and cell-cell adhesion. Typical examples are a Morse potential \citep{DOrsogna:2006ci,Middleton:2014fa}, an exponential potential \citep{Bruna:2017vr}, or a Lennard-Jones potential \citep{Jeon:2010dl}. Alternatively, the interaction between cells may be modelled as a singular hard-sphere potential
so that cells are not allowed to overlap each other, for example assuming cells
have diameter $\epsilon$ and taking $u(r)= + \infty$ for $r<\epsilon$, $u(r) =
0$ otherwise would impose that $\| \bfX_i(t) - \bfX_j(t) \| \ge \epsilon$ for all
$t$ \citep{Bruna:2012cg}. Also, cells can change type according to the following reactions:
\begin{equation}
\alpha \xrightarrow{r_\alpha} \beta, \qquad \beta \xrightarrow{r_\beta (c)} \alpha.\label{Chemo_reactions}
\end{equation}
Cells of type $\alpha$ change to type $\beta$ with constant rate
$r_\alpha$, whereas cells of type $\beta$ change type with rate $r_\beta
(c(X_\beta(t), t))$, where $X_\beta(t)$ is the position of the type $\beta$ cell
(that is, the rate may depend on the chemical concentration at the location of the
cell).

Finally, cells of type $\alpha$ secrete chemoattractant at a constant rate
$k_\alpha$, while cells of type $\beta$ consume it at a constant rate $k_\beta$.
The chemical diffuses with a diffusion constant $D_c$ and degrades with rate
$\gamma$. The chemical concentration $c(\bfx,t)$ evolves according to the PDE
\begin{align} \label{pde_c}
\partial_t c =  D_c \nabla^2 c + k_\alpha \rho_\alpha^\delta (\bfx,t) - k_\beta  \rho_\beta^\delta(\bfx,t) c - \gamma c,
\end{align}
where $\rho_\alpha^\delta$ and $\rho_\beta^\delta$ denote the random measures
for the density of cells of type $\alpha$ and $\beta$, respectively, 
\begin{equation} \label{random_measure}
	\rho_\alpha^\delta (\bfx,t) =  \sum_{i\in S_\alpha(t)} \delta(\bfx-\bfX_i(t))  , \quad \rho_\beta^\delta (\bfx,t) =  \sum_{i\in S_\beta(t)} \delta(\bfx- \bfX_i(t)). 
\end{equation}
The SDEs \eqref{eq:sde_alpha} and \eqref{eq:sde_beta} and the PDE \eqref{pde_c} are complemented with suitable initial conditions  and either periodic or no-flux boundary conditions). 
\end{subequations}

The numerical implementation of model \eqref{model} is challenging for several
reasons. In the remainder of this section we go through each of the problems one faces. In addition to the points below, it is worth keeping in mind that
generally we are interested in statistical averages of the simulations (in order
to compare them with, for example, experiments or continuum PDE models). As a
result, it is crucial that the simulation method is implemented efficiently,
exploiting parallelisation and other algorithms to speed up the simulation. 

\subsection{Time-stepping for diffusion and reactions}

The standard way to numerically integrate the SDEs \eqref{eq:sde_alpha} and
\eqref{eq:sde_beta} is to use a fixed time-step $\Delta t$ and a Euler--Maruyama
discretisation. For \eqref{eq:sde_alpha}, this reads

\begin{equation}
	\label{sde_EM}
	\bfX_i(t+ \Delta t) = \bfX_i(t) + \sqrt{2D_\alpha \Delta t} \xi_i - \sum_{j\ne i} \nabla_i u(\| \bfX_i(t) - \bfX_j(t) \|) \Delta t,
\end{equation}
where $\xi_i$ is a two-dimensional normally distributed random variable with
zero mean and unit variance. Reactions \eqref{Chemo_reactions} can also be
simulated using a fixed time-step. For example, if $N_\alpha(t)$ is the number
of $\alpha$-type cells at time $t$, then the first reaction occurs during $[t,
t+\Delta t)$ if $\zeta < N_\alpha(t) r_\alpha \Delta t$, where $\zeta$ is a
uniformly distributed random number, $\zeta \sim U(0,1)$.

The downside of the fixed time-stepping approach to simulating reactions is
that: (i) the time-step must be chosen to ensure that $N_\alpha(t) r_\alpha
\Delta t \ll 1$, which imposes a restriction on the size of $\Delta t$; (ii)
choosing such a small $\Delta t$ means that in most time-steps no reactions will
take place. Hence, lots of random numbers $\zeta$ need to be generated before
the reaction takes place \citep{Erban:2007we}; (iii) there is an exact and more
efficient simulation algorithm, the Gillespie algorithm
\citep{Gillespie:1977dc}. The Gillespie algorithm computes the time $t + \Delta t$
that the next reaction will occur as $\Delta t = \ln (1/\zeta) /(N_\alpha(t)
r_\alpha)$, where $\zeta \sim U(0,1)$ again. 

In lattice-based models for reaction-diffusion processes, space is discretised into a regular lattice and diffusion is represented as jumps between neighbouring lattices, and can be treated a reaction events. This implies that both reactions and diffusion can be implemented using the same framework, the
Gillespie algorithm being the obvious choice. In contrast, off-lattice Brownian
motion models such as \eqref{eq:sde_alpha} and \eqref{eq:sde_beta} do not fit this framework and are naturally implemented with a
fixed time-step approach \eqref{sde_EM}. As a result, the simulation algorithm
is either of fixed time-step $\Delta t$ for both processes, where $\Delta t$ is
small enough to resolve diffusion, reactions, and interactions well (see
Subsec. \ref{sec:interactions}), or fixed time-step for the cell position
updates and variable time-step for the cell number updates. 

\subsection{Cell-cell interactions} \label{sec:interactions}

Pairwise interactions between $N$ particles will generally lead to an  $O(N^2)$
loop to compute the interaction terms in \eqref{eq:sde_alpha} and
\eqref{eq:sde_beta} at every time-step $\Delta t$. If the interaction
potential $u$ is short ranged, it is convenient to use neighbourhood searches as we only need to evaluate the distances and forces between particles that are close enough. This reduces the computational cost to $O(a N)$, where $a$ is the typical number of particles in the neighbourhood of one particle.

For long-range forces, neighbourhood searches do not help as every particle
interacts significantly with every other particle. However, in many commonly
used interactions forces (for example, electrostatics, gravitational) the
interactions between well-spaced \emph{clusters} of particles can be efficiently
approximated by means of the fast multipole method (FMM)
\citep{greengard1987fast}, which also leads to a total computational cost of
$O(N)$. Aboria implements a version of the black-box FMM \citep{fong2009black},
which uses Chebyshev interpolation to approximate the interaction of
well-separated clusters. Since the present chemotaxis model uses short-range
interactions (to represent cell volume exclusion), we do not discuss the FMM
further in this paper. For more information of Aboria's FMM capabilities, the
reader is referred to the
\href{https://martinjrobins.github.io/Aboria}{documentation}.

Interactions between particles also require a careful choice of $\Delta t$ so that they are well resolved. If $\Delta t$ is too large, interactions between Brownian steps might be missed. A good rule of thumb is that the mean relative displacement (ignoring any drift terms) between two particles with diffusion coefficients $D_i$ and $D_j$ respectively, $\sqrt{2(D_i + D_j) \Delta t}$, should be less than the range of the interaction potential (equal to the sum of the particles' radii in the case of a hard-sphere interaction). It is not uncommon for short-range potentials to be singular or very steep at the origin. In these cases, the scheme \eqref{sde_EM} may not resolve well the original SDE \eqref{eq:sde_alpha} unless the time-step $\Delta t$ is prohibitively small (so that the drift term in \eqref{sde_EM} does not send particles very far apart, possibly missing other interactions on the way). An alternative to the explicit Euler--Maruyama scheme \eqref{sde_EM} is to use an implicit scheme with better convergence properties, or the so-called tamed Euler scheme \citep{Hutzenthaler:2012vi}, which modifies the drift term of \eqref{eq:sde_alpha}, $f_i (\vec X(t)) = \sum_{j\ne i} \nabla_i u(\| \bfX_i(t) - \bfX_j(t)\|) $, where $\vec X = (\bfX_1, \dots, \bfX_N)$, so that it is uniformly bounded by one:
\begin{equation}
	\label{sde_ETamed}
	\bfX_i(t+ \Delta t) = \bfX_i(t) + \sqrt{2D_\alpha \Delta t} \xi_i - \frac{f_i (\vec X(t)) \Delta t}{1 + \| f_i (\vec X(t)) \| \Delta t}.
\end{equation}
This scheme coincides with the Euler--Maruyama scheme \eqref{sde_EM} up to order
$\Delta t$ and it is just as simple to implement, but it has the advantage of
allowing larger simulation time-steps for \eqref{eq:sde_alpha} with repulsive
potentials singular at the origin. 
 
In the case of a hard-sphere interaction potential, there are several options to
implement the collisions between particles \citep[see][p. 33]{Bruna:2012ub}. One
option is to update particle positions according to \eqref{sde_EM} and correct
any overlap at time $t + \Delta t$ by moving particles apart in the direction
along the line joining the two particle centres. Namely, if the distance
between two particles is $d_{ij} = \| \bfX_i(t+\Delta t) - \bfX_j(t+\Delta t) \| <
(\epsilon_i + \epsilon_j) /2$, where $\epsilon_i$ is the diameter of the $i$th
particle, then the particles are moved apart a distance $2d_p$, where $d_p = (\epsilon_i + \epsilon_j)/2 -
d_{ij}$ is the distance that particles have penetrated
each other illegally. Note that we use twice this distance to account for the fact that particles would have collided and moved apart by $d_p$ (moving them apart  only by $d_p$ would make them be exactly in contact). The way the total update distance is distributed among
particles depends on the mean travelled distance of each particle and their diffusion coefficients. If the particles are of the same type, then
the distance is shared equally among them. If instead one particle is immobile
(suppose it is a fixed obstacle), then the total
displacement is imposed on the other particle. In general, if the particles have diffusion coefficients $D_i$ and
$D_j$, the $i$th particle takes $D_i/(D_i+D_j)$ of the displacement, and the
rest goes to particle $j$.

Another method to implement hard-sphere collisions is known as the
elastic collision method \citep{Scala:2007fd}. It consists of an event-driven
algorithm between Brownian time-steps of length $\Delta t$, whereby each
Brownian particle is attributed a ``velocity'' $V_i(t) = (\bfX_i(t+\Delta t) -
\bfX_i(t))/\Delta t$ and collisions between all particles in the interval $[t, t+
\Delta t)$ are predicted and treated using a standard event-driven method for
ballistic dynamics. On the one hand, this method predicts rather than corrects
collisions, and it is therefore more accurate than the first one. This implies
that one may take larger steps $\Delta t$. On the other hand, the event-driven
method is computationally more intensive and is more complex to implement. 
 
\subsection{Spatial matching between discrete and continuous variables}

The hybrid model \eqref{model} combines two modelling frameworks: a
particle-based approach for the cells, equations \eqref{eq:sde_alpha} and
\eqref{eq:sde_beta}, and a continuum approach for the chemical concentration,
\eqref{pde_c}. This means that we have to consider each part of the model
separately and establish a way to match them in space.

One approach is that taken by \cite{Newman:2004cj}, where the chemical
concentration $c(\bfx,t)$ is found by formally integrating equation \eqref{pde_c}
along the cell paths. Assuming $\Omega = \mathbb R^2$, the result is 
\begin{multline} \label{chem_expl}
	c(\bfx,t) = \int_0^t \int_{\mathbb R^2}  G_\gamma(\bfx-\bfx', t-t') \left[ k_\alpha  \rho_\alpha^\delta(\bfx',t')  - k_\beta  \rho_\beta^\delta(\bfx',t') \right] \ud x' \ud t',
\end{multline}
where $\rho_\alpha^\delta$ and $\rho_\alpha^\delta$ are given in \eqref{random_measure} and $G_\gamma$ is the Green's function for the chemical diffusion equation in $\mathbb R^2$,
\begin{equation}
	G_\gamma(\bfx,t) = (4 \pi D_c t)^{-1} \exp\left ( -\frac{\| \bfx\|^2}{4D_c t} - \gamma t \right).
\end{equation}
Since we have an explicit expression for the chemical concentration in the whole
space, the chemical gradient in \eqref{eq:sde_beta} and the chemical
concentration in the reaction rate in \eqref{Chemo_reactions} can be evaluated
exactly. In particular, the gradient in \eqref{eq:sde_beta} can be written as
\begin{multline}
	\nabla c(\bfX_i(t),t) = \int_0^t \Big[  \sum_{j \in S_\alpha(t')} k_\alpha \nabla G_\gamma (\bfX_i(t) - \bfX_j(t'),t-t') \\ - \sum_{j \in S_\beta(t')}  k_\beta\nabla G_\gamma (\bfX_i(t) - \bfX_j(t'),t-t') \Big] \ud t'.
	\label{eq:sde_beta_green} 
\end{multline}
Therefore, this approach requires the history of the cell positions $\bfX_i(t)$ only, and no explicit evaluation of the chemical concentration (unless $c$ is required as a simulation output, in which case one uses \eqref{chem_expl}). 
One major drawback of this approach is that it  only works when the domain $\Omega$ is the whole space, and therefore it may not be applicable in many cases. 

The alternative approach is to discretise and integrate the chemical
concentration on a grid, for example using a finite-differences method. This approach works for bounded domains, but it has the disadvantage that it requires spatial
matching between the discrete and continuum variables. Specifically, the
particles can be positioned at an arbitrary point inside the domain, while the
chemical concentration is only calculated at grid points $\bfp_1, \dots , \bfp_L \in
\Omega$. Then we have a two-way matching to do: interpolate the concentration at
the off-grid particle positions to simulate \eqref{eq:sde_beta} and
\eqref{Chemo_reactions}, and generate estimates for the cell densities
$\rho_\alpha^\delta$ and $\rho_\beta^\delta$ at the points $\bfp_l$ to integrate
\eqref{pde_c}. 

The approximation of $c(\bfx,t)$ and $\nabla c(\bfx,t)$ at points $\bfX_i(t)$ can be done
using a variety of interpolation methods, such as linear or spline
interpolation. Since the grid points $\bfp_l$ on which $c$ is computed are chosen
beforehand to provide a good approximation of $c$, they will generally also form
a good set of interpolating points \citep{Franz:2013ds}.

The interpolation of the cell densities $\rho_\alpha^\delta$ and
$\rho_\beta^\delta$ from the cell positions $\bfX_i$ to the grid points $\bfp_l$,
necessary to update $c(\bfp_l,t)$ according to \eqref{pde_c}, is slightly more
delicate. A basic approach would be to ``shift'' the delta function from $\bfX_i$ to
its closest grid point $\bfp_l$, so that $\rho_\alpha^\delta(\bfp_l,t)$ is a count of the
number of $\alpha$-type cells in the neighbourhood of $\bfp_l$. However, this is
quite a crude approximation to make, rendering the more accurate approaches in
the discretisation of the equations or the interpolation of $c$ a waste of effort.
This is why the standard approach is to obtain a continuous approximation of the density, or density estimate, and
then evaluate it at the grid points $\bfp_l$. One way to achieve this is to use a
particle-in-cell method with piecewise linear polynomials as done by
\cite{Dallon:1997be}. They use a square lattice and the mass of a
delta function at $\bfX_i$ is distributed among the four nearest grid points proportionally to their distances. Another way is to use a kernel density
estimation, which is generally used to estimate the probability density of a
random process from a large number of iterations \citep[for details see][]{Franz:2013ds}. The estimate of the density of type $\alpha$ cells is found as
\begin{equation}
	\rho_\alpha (\bfx,t) =  \sum_{i\in S_\alpha(t)} K_h(\bfx-\bfX_i(t)) = K_h(\bfx) \ast \rho_\alpha^\delta (\bfx,t),
\end{equation}
where $\ast$ is the spatial convolution, and $K_h(\bfx)$ denotes a kernel of bandwidth $h$, $K_h(\bfx) = h^{-2}K(\bfx/h)$, taken to be a continuous, symmetric and normalised function. The idea is that, as the bandwidth parameter $h \to 0$, the estimate $\rho_\alpha$ approximates the sum of delta functions in $\rho_\alpha^\delta$. The Gaussian kernel, $K(\bfx) = (2 \pi)^{-1} \exp(-\| \bfx\|^2/2)$, is one of the most commonly used kernels for density estimation. In the context of hybrid modelling, the Gaussian kernel density estimation was used in \cite{Franz:2013do,McLennan:2012jb}. 

The earliest hybrid models for chemotaxis modelled cells as point particles.
Accordingly, the formulation of \eqref{pde_c} with Dirac deltas was considered
to be the exact model, and the kernel density estimation its approximation.
However, if the actual size of cells is taken into account, it makes sense to
assume that the cells consume or degrade chemical all along their shape, and not
only in the centre. In this case, the bandwidth parameter $h$ can be thought of
the lengthscale of the cell (for example, $h$ can be related to the hard-sphere
diameter when cells are modelled as hard bodies), independent of the grid
spacing for the chemical \citep{McLennan:2012jb}.

\section{Model implementation with Aboria}
\label{sec:Implementation}

\lstMakeShortInline[basicstyle=\ttfamily,keywordstyle=\color{blue}\ttfamily]`
\lstset{language=C++,
	columns=flexible,
    frame = single,
    basicstyle=\small\ttfamily,
    keywordstyle=\small\color{blue}\ttfamily,
    stringstyle=\small\color{red}\ttfamily,
    commentstyle=\small\color{magenta}\ttfamily,
    morecomment=\small[l][\color{green}]{\#}
}

We consider an example in a square domain, $\Omega =
[-L/2,L/2]^2$, with no-flux boundary conditions. Cells interact with each other
via a soft short-range repulsive potential $u(r) =  \exp(-r/\epsilon)$, with
$\epsilon = 0.01$.  Initially there are the same number of  particles of each type, $N_\alpha =
N_\beta$, and $N = N_\alpha + N_\beta$. Cells of type $\alpha$ are distributed according to a two-dimensional
normal centred at the origin and $\sigma = 0.1$. Cells of type $\beta$ are
uniformly distributed in the whole domain. Initially there is no chemical in the domain. 

\subsection{Particle set with two types of particles}

We define three Aboria variables: `type` to refer to the cell type (`type=true`
for cells of type $\alpha$, and `type=false` for cells of type $\beta$),
`concentration` for the concentration of $c$ at the location of the particle,
and `drift` to store the two-dimensional  drift vector $\chi \nabla c$ (using
an Aboria `vdouble2` type, representing a two-dimensional vector). We then
define the particle set type, given by `Particles_t`, which contains the Aboria
variables and has a spatial dimension of two (specified by the second template
argument). For convenience, we also define `position` as the
`Particles_t::position` subclass (we will use this later on):

\noindent\begin{minipage}{\linewidth}
\begin{lstlisting}[language=c++,firstline=15, lastline=22]
    ABORIA_VARIABLE(conc, double, "conc");
    ABORIA_VARIABLE(drift, vdouble2, "drift");
    ABORIA_VARIABLE(starting, vdouble2, "starting_position");
    ABORIA_VARIABLE(next_position, vdouble2, "next position");
    ABORIA_VARIABLE(type, uint8_t, "type");
    typedef Particles<std::tuple<type, drift, conc, starting, next_position>, 2>
        Particles_t;
    typedef typename Particles_t::position position;
\end{lstlisting}
\end{minipage}

Finally we create `particles`, an instance of `Particles_t`, containing $N$
particles, and initialise the random seed of the set based on a unique sample
number `sample`:

\noindent\begin{minipage}{\linewidth}
\begin{lstlisting}
Particles_t particles(N);
particles.set_seed(N * sample);
\end{lstlisting}
\end{minipage}

In Aboria there are two ways to access and operate on `particles`, either in
low-level language or a high-level symbolic language. We show how these two
approaches work when initialising the positions of the $N$ particles. The
low-level approach uses the standard C\texttt{++} random library to generate the
gaussian and normal distributions, and loops over the particles to set their
positions. We define `min` and `max` as vectors representing respectively the lower and upper
boundaries for each dimension, `min` $=(-L/2, -L/2)$ and `max`$=(L/2, L/2)$:

\noindent\begin{minipage}{\linewidth}
\begin{lstlisting}[language=c++,firstline=78, lastline=88]
    std::uniform_real_distribution<double> uniform(min[0], max[0]);
    std::normal_distribution<double> normal(0, 0.1 * (max[0] - min[0]));
    for (size_t i = 0; i < N; ++i) {
        get<type>(particles)[i] = i < Na;
        auto &gen = get<generator>(particles)[i];
        if (get<type>(particles)[i]) {
            get<position>(particles)[i] = vdouble2(normal(gen), normal(gen));
        } else {
            get<position>(particles)[i] = vdouble2(uniform(gen), uniform(gen));
        }
    }
\end{lstlisting}
\end{minipage}

For the high-level symbolic approach, we define symbolic objects `x` and `typ`,
representing the `position` and `type` variables (and others that we will use
later on).  We also define two Aboria random number
generators; `normal` for normally distributed numbers, and `uni` for uniformly
distributed numbers. Finally, we also create a label object `k` associated to the particle set.
This label performs a similar function to the $i$ subscript for the variable
$\bfX_i(t)$ in Eq.~\ref{sde_EM}. All operations involving `k` are implicitly
performed over the entire particle set.

\noindent\begin{minipage}{\linewidth}
\begin{lstlisting}
  Symbol<position> x;
  Symbol<type> typ;
  VectorSymbolic<double, 2> vector;
  Normal normal;
  Uniform uni;
  Label<0, Particles_t> k(particles);
\end{lstlisting}
\end{minipage}

Then the high-level initialisation of the positions of `particles` (assuming
that the `type` variable has already been set) is:

\noindent\begin{minipage}{\linewidth}
\begin{lstlisting}
const double sigma = 0.1 * (max[0] - min[0]);
const double width = max[0]-min[0];
x[k] = if_else(typ[k]
        , sigma * vector(normal[k], normal[k])
        , width * vector(uni[k], uni[k]) + min); 
\end{lstlisting}
\end{minipage}

The advantage of the high-level approach is that we can directly write
expressions which are meant to be evaluated over the whole particle set.
However, as it implements a custom eDSL, it is by definition limited to
operations that can be expressed by this language. For example, the interaction
between the chemical grid and the individual particles cannot be expressed with
this DSL. Therefore, for the remainder of this paper we will proceed by
implementing the model using the lower level interface. 

Finally, we initialise the spatial search data structure, providing it with
lower (`min`) and upper (`max`) bounds for the domain, and setting non-periodic
boundary conditions (`periodic` in this case is `false`):

\noindent\begin{minipage}{\linewidth}
\begin{lstlisting}[language=c++,firstline=104, lastline=104,label={lst:init_neighbour_search}]
    particles.init_neighbour_search(min, max, vbool2::Constant(periodic));
\end{lstlisting}
\end{minipage}

This subdivides the computational domain into square cells of equal size. The
default cell side length is such that, if particles were uniformly distributed
in the domain, then there would be on average ten particles per cell. However, it is
also possible to specify a desired side length as a forth parameter (to make it
equal, for example, to a cut-off distance for the calculation of interaction
forces). 

\subsection{Equation of motion of cells} 

Here we explain how to implement the SDEs \eqref{eq:sde_alpha} and
\eqref{eq:sde_beta}. For simplicity, in this section we are going to assume that
the chemical drift $\chi\nabla c$ in \eqref{eq:sde_beta} is fixed and already
stored in the `drift` variable within each $\beta$-particle. To implement the
interaction force between two particles $\bfX_i$ and $\bfX_j$, we use the
`euclidean_search` function in Aboria, which returns an iterator `j` that
iterates through all the particles within a certain radius (`cutoff`) of a
given point. For each potential pair of particles, the iterator `j` can provide
the shortest vector between $i$ and $j$ (normally $\bfX_j-\bfX_i$, but not always 
for periodic simulations), and we use this vector to evaluate the interaction
force between $i$ and $j$:

\noindent\begin{minipage}{\linewidth}
\begin{lstlisting}[language=c++,firstline=170, lastline=179]
            for (auto j = euclidean_search(particles.get_query(),
                                           get<position>(i), cutoff);
                 j != false; ++j) {
                const double r = j.dx().norm();
                if (r > 0.0) {
                    get<next_position>(i) += -inter * (dt / epsilon) *
                                             std::exp(-r / epsilon) *
                                             (j.dx() / r);
                }
            }
\end{lstlisting}
\end{minipage}

Note that we increment the variable `next_position`, rather than `position`, as
this interaction loop uses the particle positions and therefore cannot update
them until the loop is complete.

Now we can implement both the drift and the Brownian diffusion of a particle using
the stored `drift` variable, as well as a random number generator that is stored
in the variable `generator`. We will use the standard C++ normal distribution
(the `normal` variable) to generate two normally distributed random numbers used
for the diffusion:

\noindent\begin{minipage}{\linewidth}
\begin{lstlisting}[language=c++,firstline=183, lastline=187]
            auto &gen = get<generator>(i);
            const auto D = get<type>(i) ? Da : Db;
            get<next_position>(i) +=
                dt * get<drift>(i) +
                std::sqrt(2 * D * dt) * vdouble2(normal(gen), normal(gen));
\end{lstlisting}
\end{minipage}

The no-flux boundary conditions are enforced by reflections if the particles end
up outside the domain:

\noindent\begin{minipage}{\linewidth}
\begin{lstlisting}[language=c++,firstline=198, lastline=205,label={lst:no-flux}]
                for (size_t d = 0; d < 2; ++d) {
                    if (get<next_position>(i)[d] < min[d]) {
                        get<next_position>(i)[d] =
                            -L - get<next_position>(i)[d];
                    } else if (get<next_position>(i)[d] > max[d]) {
                        get<next_position>(i)[d] = L - get<next_position>(i)[d];
                    }
                }
\end{lstlisting}
\end{minipage}

\subsection{Hard-sphere interactions}

In the previous subsection, we showed how to implement soft interactions between
particles. Below we show an example of the update of `x` if,  instead of using an
interaction potential, we want to model cells as hard spheres of diameter
`epsilon`. Using the first of the two hard-sphere collision algorithms
discussed, the new particle positions are obtained using: 

\noindent\begin{minipage}{\linewidth}
\begin{lstlisting}
    for (auto j = euclidean_search(particles.get_query(), get<position>(i), 
          epsilon); j != false; ++j) {
      const double r = j.dx().norm();
      if (r > 0.0) {
        const double D = get<type>(i) ? Da : Db;
        get<next_position>(i) +=
            -(2 * D / (Da + Db)) * (epsilon/r - 1) * j.dx();
      }
    }
\end{lstlisting}
\end{minipage}

\subsection{Reactions between cells}

The reactions \eqref{Chemo_reactions} to change type between cells are
implemented as follows:

\noindent\begin{minipage}{\linewidth}
\begin{lstlisting}[language=c++,firstline=191, lastline=193]
            const double reaction_propensity =
                (get<type>(i) ? ra : rb * get<conc>(i)) * dt;
            get<type>(i) ^= uniform(gen) < reaction_propensity;
\end{lstlisting}
\end{minipage}

If the `type` variable is `true` ($\alpha$ cell), then the cell changes type if $u <
r_\alpha \Delta t$, where $u \sim U(0,1)$. If instead `type` is `false` ($\beta$
cell), then the reaction takes place if $u < r_\beta  c(\bfX_i(t),t) \Delta t$. In our
implementation, the chemical concentration at the position of the $i$th particle
is saved in the `conc` variable (we explain in the next section how this is
updated). 

\subsection{Chemical concentration field}

The chemical is modelled by its continuum concentration rather than individual
particles. For this reason, instead of using an Aboria `Particles_t` set to
describe it, we use a standard C\texttt{++} linear algebra library. We take the
computational domain for the chemical to be equal to that of the cells, that is,
$\Omega$. We discretise the domain in $N_c$ grid points, $\bfp_l$, in each
direction, and integrate the PDE \eqref{pde_c} for the chemical using finite
differences (second-order centred differences in space, and forward Euler in
time). `Vector` is a vector type of length $N_c^2$ used to create instances `c`
for the discretised chemical concentration $c$, `rhoalpha` for the $\alpha$-type
cell density estimate $\rho_\alpha$, and `rhobeta` for the $\beta$-type cell
density estimate $\rho_\beta$. `Vector2` is a matrix type of size $N_c^2 \times
2$ to store the gradient of the chemical concentration $\nabla c$. Finally,
`SparseMatrix` is a sparse matrix type of size $N_c^2 \times N_c^2$ that is used
to store the various finite differences discretisation matrices: `D2`
stores $I - \Delta t D^2_{xy}$, where $D^2_{xy}$ is the discrete Laplacian and
$\Delta t$ is the time-step,  and `D1x` and `D1y` that perform the first
derivatives with respect to the horizontal and vertical coordinates,
respectively:

\noindent\begin{minipage}{\linewidth}
\begin{lstlisting}[language=c++,firstline=31, lastline=37]
    Vector c;
    Vector rhoalpha;
    Vector rhobeta;
    Vector2 grad_c;
    SparseMatrix D2;
    SparseMatrix D1x;
    SparseMatrix D1y;
\end{lstlisting}
\end{minipage}

Then we define the Gaussian kernel $K_\epsilon$ to estimate the cell densities
$\rho_\alpha$ and $\rho_\beta$, rescaled with $\epsilon$ (the cell diameter)
since we assume cells produce/consume chemical using receptors that are
distributed over their bodies \citep{McLennan:2012jb}:

\noindent\begin{minipage}{\linewidth}
\begin{lstlisting}[language=c++,firstline=118, lastline=122]
    const double kernel_scale1 = 1.0 / (2.0 * PI * std::pow(epsilon, 2));
    const double kernel_scale2 = 1.0 / (2.0 * std::pow(epsilon, 2));
    auto Kbw = [&](const vdouble2 &dx) {
        return kernel_scale1 * std::exp(-dx.squaredNorm() * kernel_scale2);
    };
\end{lstlisting}
\end{minipage}

We approximate the chemical concentration and the cell density estimates at
the grid points.  The position of the grid point $\bfp_0$ closest to the left
bottom corner of the computational domain is $\bfp_0 = (-L/2, -L/2)$ and, as mentioned above, is stored
as `min`. The code below computes the discretised cell density estimates, finding
the cells of a given type that are within a `cutoff` distance of the grid point
$\bfp_l$ where we want to compute `rhoalpha` and `rhobeta` at. Then we evaluate the
kernel $K_\epsilon$ at $\bfp_l - \bfX_i(t)$ and add it to the corresponding vector and
grid position:

\noindent\begin{minipage}{\linewidth}
\begin{lstlisting}[language=c++,firstline=228, lastline=244]
        rhoalpha.setZero();
        rhobeta.setZero();
        for (int k = 0; k < Nc; ++k) {
            for (int l = 0; l < Nc; ++l) {
                const vdouble2 rgrid = min + vint2(k, l) * hc;
                const int ind_linear = k * Nc + l;
                for (auto i =
                         euclidean_search(particles.get_query(), rgrid, cutoff);
                     i != false; ++i) {
                    if (get<type>(*i)) {
                        rhoalpha(ind_linear) += Kbw(i.dx());
                    } else {
                        rhobeta(ind_linear) += Kbw(i.dx());
                    }
                }
            }
        }
\end{lstlisting}
\end{minipage}

We implement \eqref{pde_c} to update the chemical concentration vector `c` using:

\noindent\begin{minipage}{\linewidth}
\begin{lstlisting}[language=c++,firstline=248, lastline=249]
        c = D2 * c +
            dt * (ka * rhoalpha - kb * c.cwiseProduct(rhobeta) - gam * c);
\end{lstlisting}
\end{minipage}

Then we calculate the concentration gradient `grad_c` like so:

\noindent\begin{minipage}{\linewidth}
\begin{lstlisting}[language=c++,firstline=129, lastline=130]
        grad_c.col(0) = chi * D1x * c;  // drift in x-direction
        grad_c.col(1) = chi * D1y * c;  // drift in y-direction
\end{lstlisting}
\end{minipage}

\subsection{Spatial matching from regular grid for the chemical to particle positions}

Finally, we need to convert the continuum variables $c$ and $\nabla c$ (which in
the numerical simulation are approximated at regular grid points) to the
positions of $\beta$-type cells, so that we can evaluate the drift in
\eqref{eq:sde_beta} and the reaction rate $r_\beta(c)$ in
\eqref{Chemo_reactions}. To do that, we use linear interpolation:

\noindent\begin{minipage}{\linewidth}
\begin{lstlisting}[language=c++,firstline=140, lastline=166]
            if (get<type>(i)) {
                get<drift>(i) = vdouble2(0, 0);
                get<conc>(i) = 0;
            } else {
                auto &x = get<position>(i);

                const vint2 ind = Aboria::floor((x - min) / hc);

                // linear index to access fx, fy
                const int ind_linear = ind[0] * Nc + ind[1];

                // interpolate gradient
                const vdouble2 x_low = ind * hc + min;
                get<drift>(i) = vdouble2(
                    grad_c(ind_linear, 0) +
                        (grad_c(ind_linear + Nc, 0) - grad_c(ind_linear, 0)) *
                            (x[0] - x_low[0]) * Nc,
                    grad_c(ind_linear, 1) +
                        (grad_c(ind_linear + 1, 1) - grad_c(ind_linear, 1)) *
                            (x[1] - x_low[1]) * Nc);

                get<conc>(i) = c[ind_linear] +
                               (c[ind_linear + Nc] - c[ind_linear]) *
                                   (x[0] - x_low[0]) * Nc +
                               (c[ind_linear + 1] - c[ind_linear]) *
                                   (x[1] - x_low[1]) * Nc;
            }
\end{lstlisting}
\end{minipage}

\section{Simulation results: output and analysis of data}
\label{sec:Results}

While the C\texttt{++} language is ideal for implementing fast simulations with
low memory overhead, the Python language is generally preferred for plotting, and 
pre- and post-processing of simulation data. Thankfully, there are many tools
that enable wrapping of C\texttt{++} code in Python, and we will make use of one
of these,
\href{http://www.boost.org/doc/libs/1_66_0/libs/python/doc/html/article.html}{Boost
Python}, to enable us to call our simulation code from Python.

The main difficulty in wrapping C\texttt{++} code in Python is transferring data
between the two languages. In our case, we only need to transfer data to Python
for post-processing and plotting. We can use the
\href{https://www.boost.org/doc/libs/1_67_0/libs/python/doc/html/numpy/index.html}{Boost
NumPy} extension to wrap a Numpy array around a Aboria variable given in the
template argument `V`. 

\noindent\begin{minipage}{\linewidth}
\begin{lstlisting}[language=c++,firstline=188, lastline=199]
    template <typename V>
    p::object get_particle_vector() {
        using data_t = typename V::value_type::value_type;
        const size_t N = V::value_type::size;
        np::dtype dt = np::dtype::get_builtin<data_t>();
        p::tuple shape = p::make_tuple(particles.size(), N);
        p::tuple stride = p::make_tuple(sizeof(data_t) * N, sizeof(data_t));
        p::object own;
        return np::from_data(
            reinterpret_cast<double *>(get<V>(particles).data()), dt, shape,
            stride, own);
    }
\end{lstlisting}
\end{minipage}

Note that no copying of data occurs in this function, the new Numpy array that
is returned from the function simply wraps the data so that it can be easily
accessed in Python. This function assumes that the Aboria variable is a vector
type (e.g. `position`), but we can easily write another function to wrap a
scalar variable (e.g. `type`). Note also that we can write a very similar
function to transfer the grid data to Python as well (see the full code in the
Supplementary Material for all three of these functions).

Now that we can transfer data, we need a C\texttt{++} object with which we can
interact in Python. Thus we will create a `Simulation` class to store our
data, with functions like `integrate()` and `get_positions()` that will allow us
to either integrate the simulation forward in time, or obtain internal variables
for plotting:

\noindent\begin{minipage}{\linewidth}
\begin{lstlisting}
class Simulation {
  const double PI = boost::math::constants::pi<double>();
  const double epsilon = 0.01; // interaction range
  // other simulation constants go here....

  Particles_t particles;
  Vector c;
  // other data objects go here....

public:
  // create a Simulation object, with a `sample` seed to
  // initialise the random number generator
  Simulation(const size_t sample);

  // integrate the simulation forward in time by `time`
  void integrate(const double time);

  // return the particle positions as a numpy array
  p::object get_positions() {
    return get_particle_vector<position>();
  }

  // other data access functions go here...

};

\end{lstlisting}
\end{minipage}

Now we can use
\href{http://www.boost.org/doc/libs/1_66_0/libs/python/doc/html/article.html}{Boost
Python} to wrap our `Simulation` class and enable it to be used from Python: 

\noindent\begin{minipage}{\linewidth}
\begin{lstlisting}
using namespace boost::python;

BOOST_PYTHON_MODULE(chemo) {
  numpy::initialize();
  class_<Simulation>("Simulation", init<size_t, int>())
      .def("integrate", &Simulation::integrate)
      .def("get_positions", &Simulation::get_positions)
      // other data access functions here...
      ;
}
\end{lstlisting}
\end{minipage}

After this we need to compile the code we gave generated thus far. For the example code
included with this paper we have used the CMake build system (see the 
`CMakeLists.txt` file included in the Supplementary Material for details of how 
this is done). After compilation we are left with a final library file named 
`chemo` (with an extension that depends on the particular operating systems you 
are using) that we can use in Python like so:

\noindent\begin{minipage}{\linewidth}
\begin{lstlisting}[language=Python]
import chemo
import matplotlib.pyplot as plt

sim = chemo.Simulation(1)
sim.integrate(0.05)
positions = sim.get_positions()
the_type = sim.get_type()
plt.scatter(positions[:, 0], positions[:, 1],  c=the_type, lw=0)
plt.show()
\end{lstlisting}
\end{minipage}

The above listing simple integrates the simulation until $T_f=0.05$, and then
creates a scatter plot of the cells coloured by their type. This simulation
consists of a single sample, and creating new simulation objects with a
different initial seed (in the code above the sample seed is set to 1) will
result in a different random realisation.

In order to gain a complete picture of
the dynamics we need to run multiple samples, and average the results. However,
now that our simulation code is running in Python, we can use its high-level
features and libraries to do this relatively easily. For example, we can use
Python's standard
\href{https://docs.python.org/2/library/multiprocessing.html}{multiprocessing}
 library to run many simulations in parallel, average the
particle positions by calculating histograms using Numpy's
\href{https://docs.scipy.org/doc/numpy/reference/generated/numpy.histogram2d.html}{histogram2d}
function, cache the results of the simulations to disk using Python
\href{https://docs.python.org/2/library/pickle.html}{pickle}, and finally plot the
results using \href{https://matplotlib.org/}{matplotlib}. We will not
explain these facilities in detail here, but instead refer readers to the external
documentation links provided, and to the Supplementary Material for a full code
listing in `paper_plots.py` showing how this might be done. 

\begin{figure}[htb]
  \centering
  \includegraphics[width=1.0\textwidth]{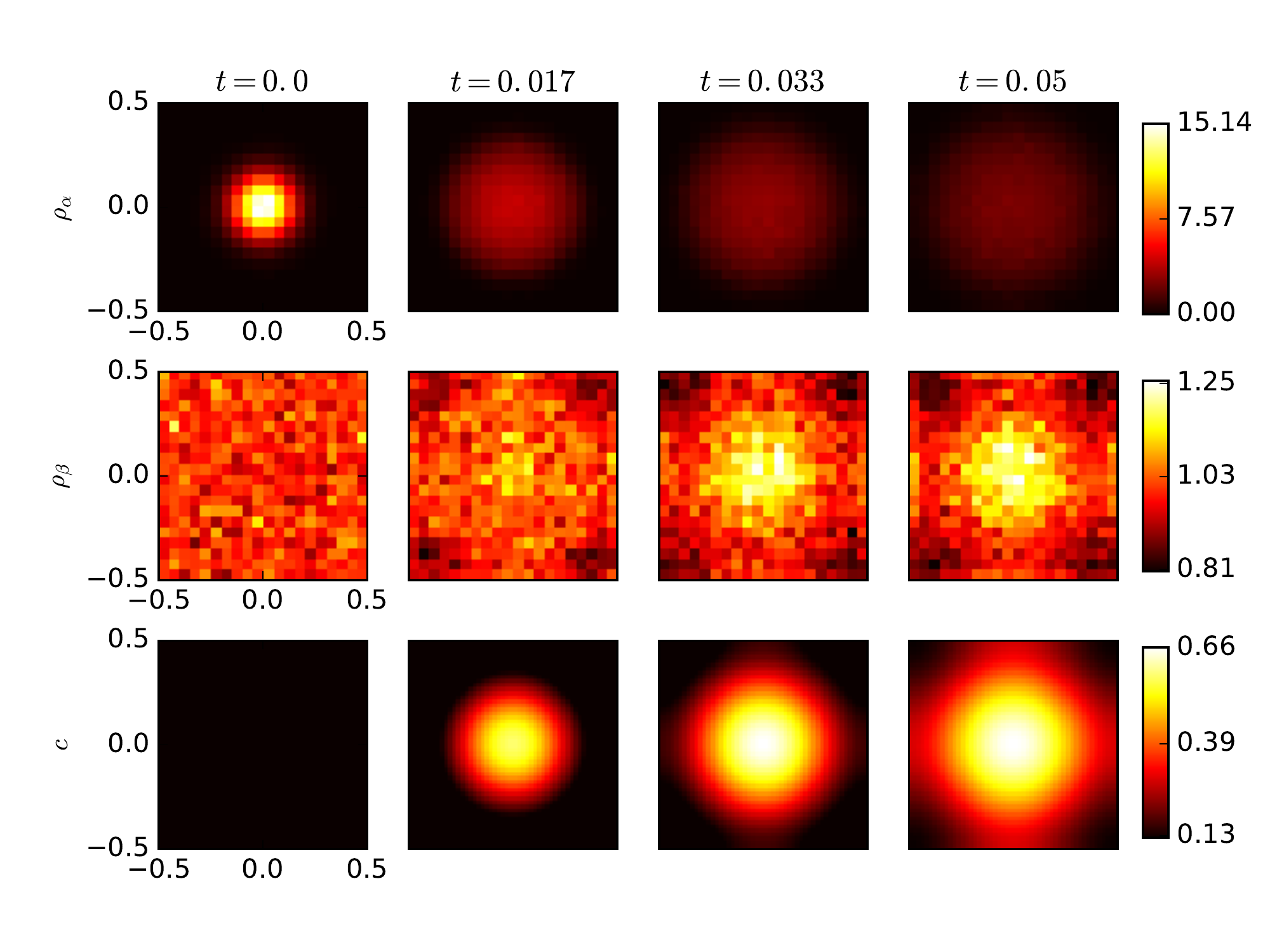}
      \caption{\label{fig:averagedVis} Visualisation of the simulation at times
      $t=0, T_f/3, 2 T_f/3$ and $t = T_f = 0.05$, averaged over 2000 random
      realisations. (Top row) Histograms of the $\alpha$ particle density
      $\rho_\alpha$ normalised by $N_{\alpha}$. (Middle row) Histograms of the
      $\beta$ particle density $\rho_\beta$ normalised by $N_{\beta}$. (Bottom
      row) Chemical concentration $c$. Parameters used: $N_\alpha(0) =
      N_\beta(0) = 100, \epsilon = 0.02, r_\alpha = 10, r_\beta = 0, D_\alpha =
      0.1, D_\beta = 1,D_c = 1, k_\alpha=0.1, k_\beta=0.03, \Delta t =
      \frac{(0.23 \epsilon)^2}{4 D_\beta}, N_c=52$.}
\end{figure}

The generated figures from `paper_plots.py` are plotted in Figure
\ref{fig:averagedVis}, showing the averaged histograms for the $\alpha$ and
$\beta$ particle types, as well as the averaged distribution of the chemical
concentration $c$. The upper row shows the diffusion of $\alpha$ particles from
an initial Gaussian profile becoming increasingly spread out over the domain.  
The middle row shows the diffusion of $\beta$ particles from an initial uniform 
gradient. As the simulation proceeds the $\alpha$ particles produce the chemical 
in the centre of the domain, and soon the chemotactic gradient term results in 
$\beta$ particles becoming clustered in the middle of the domain. The influence 
of the no-flux boundary conditions is seen as a build-up of $\beta$ particles 
near all four boundaries. 

\begin{figure}[htb]
  \centering
  \includegraphics[width=0.7\textwidth]{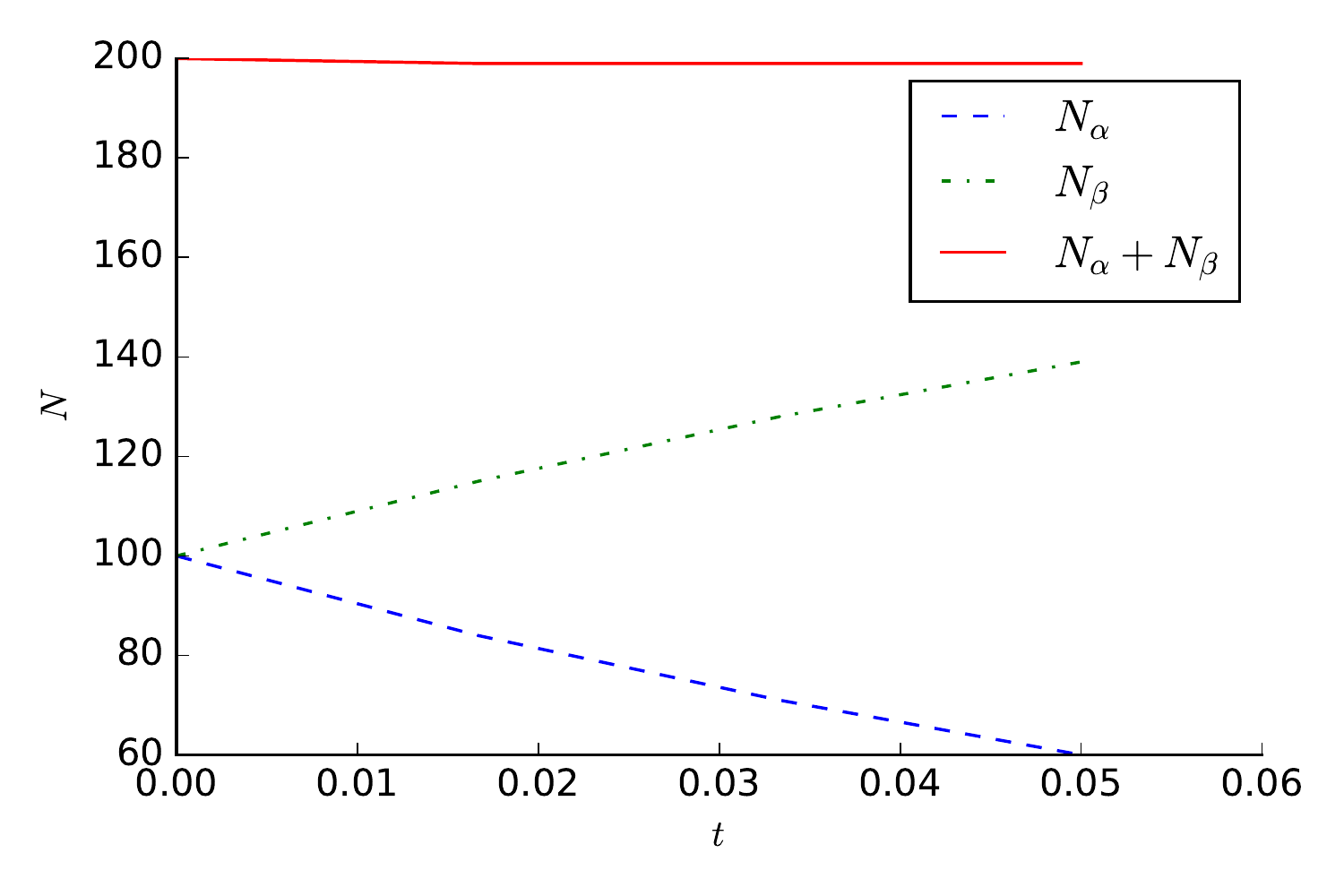}
      \caption{\label{fig:nparticles} The evolution in time of the number of
      particles in each species, $N_\alpha$ and $N_\beta$, as well as the total
      number of particles $N_\alpha + N_\beta$. Results averaged over 2000
      random realisations of the simulation.}
\end{figure}

Figure \ref{fig:nparticles} shows the number of $\alpha$ and $\beta$
particles, as well as the total number of all particles. This shows the dominant
conversion of $\beta$ particles to $\alpha$, as well as the conservation of
$N_{\alpha} + N_{\beta}$ over time.

Once we have the simulation framework established, we can begin conducting
numerical experiments, altering the domain, parameters, initial distributions or
behaviours of particles in order to explore different chemotactic behaviours, or
taking advantage of the transparency of numerical simulation by measuring
different quantities of interest. For example, in order to explore the average
mean squared displacement for the $\alpha$ and $\beta$ particles we might wish
to run another experiment using periodic boundary conditions, and track the
total displacement of an individual particle from its initial position. We can
do this by defining a new variable `starting`:

\noindent\begin{minipage}{\linewidth}
\begin{lstlisting}[language=c++,firstline=17, lastline=17]
    ABORIA_VARIABLE(starting, vdouble2, "starting_position");
\end{lstlisting}
\end{minipage}

\noindent and then initialising it to the starting position of each particle at the
beginning of the simulation.

\noindent\begin{minipage}{\linewidth}
\begin{lstlisting}[language=c++,firstline=98, lastline=100]
    for (size_t i = 0; i < N; ++i) {
        get<starting>(particles)[i] = get<position>(particles)[i];
    }
\end{lstlisting}
\end{minipage}

Instead of the no-flux boundary conditions that we implemented previously (in
Section \ref{lst:no-flux}), we will use periodic boundary conditions.  Note that
setting the periodic argument of `init_neighbour_search` to `true` (see Section
\ref{lst:init_neighbour_search}) will cause particles that cross the periodic
boundary to be automatically moved to the opposite boundary. We will counteract
this by updating `starting` whenever a particle will cross the periodic
boundary, so that the particle displacement thus far is preserved.

\noindent\begin{minipage}{\linewidth}
\begin{lstlisting}[language=c++,firstline=209, lastline=215]
                for (size_t d = 0; d < 2; ++d) {
                    if (get<next_position>(i)[d] < min[d]) {
                        get<starting>(i)[d] += L;
                    } else if (get<next_position>(i)[d] >= max[d]) {
                        get<starting>(i)[d] -= L;
                    }
                }
\end{lstlisting}
\end{minipage}

Once we have our new variable in place we can simply calculate the mean squared
displacement (MSD) of each particle by comparing its current `position` to the
`starting` variable. We can use our Python wrapper to obtain both of these
variables and calculate the MSD as a post-processing step.

\begin{figure}[htb]
  \centering
  \includegraphics[width=0.7\textwidth]{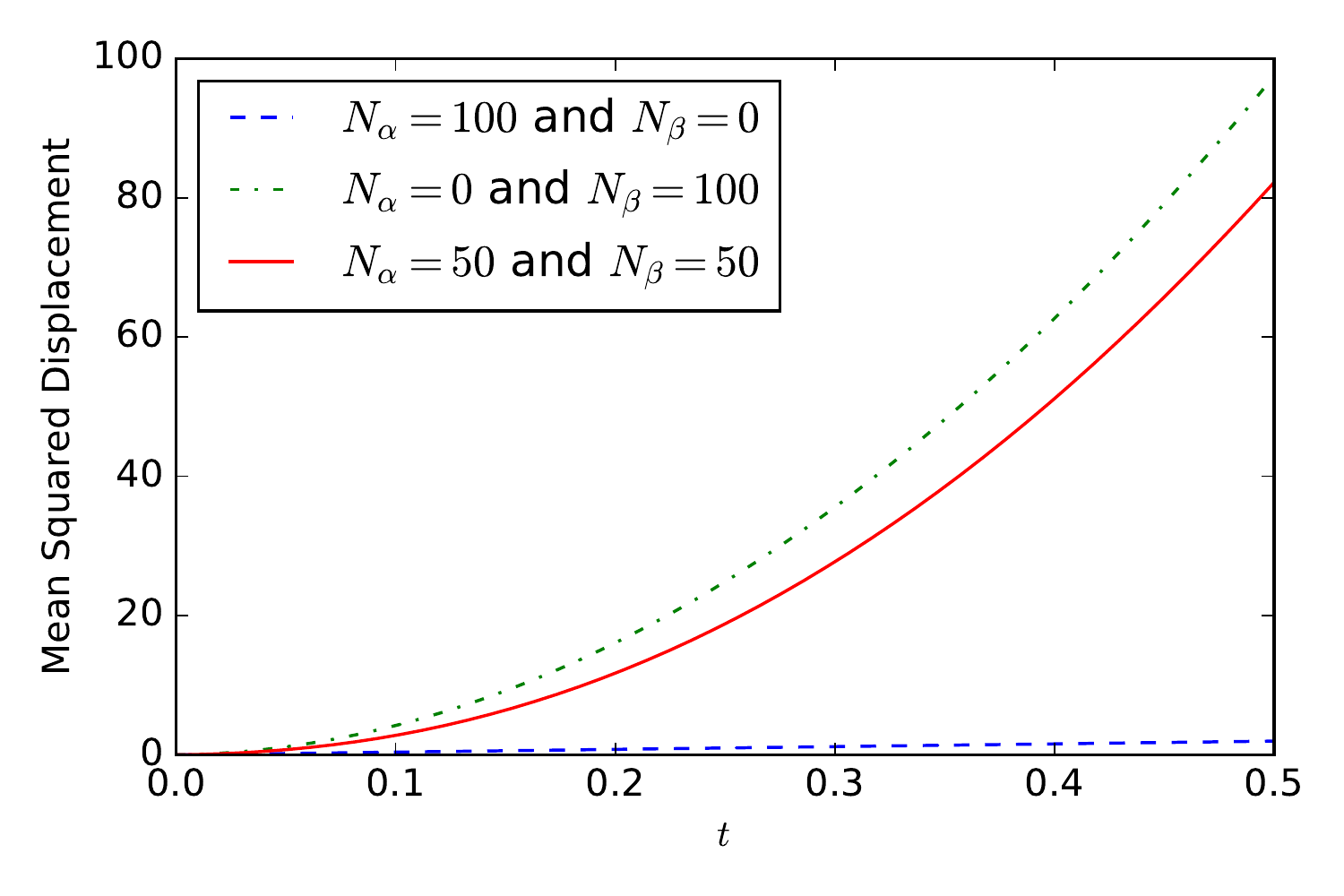}
      \caption{\label{fig:msd} The Mean Squared Displacement (MSD) was
calculated for three different simulations: (1) $N_\alpha=100, N_\beta=0$ with
reactions off, (2) $N_\alpha=0, N_\beta=100$ with reactions off, (3)
$N_\alpha=50,N_\beta=50$, with reactions on. In all cases the chemical
concentration was set to $c(\bfx) = x$ to induce a constant drift term for the
$\beta$ particles. The MSD for (1) is purely diffusion driven and shows the
correct linear growth. The MSD for (2) and (3) is dominated by the drift term on
the $\beta$ particles, which gives the $t^2$ growth.}
\end{figure}

Figure \ref{fig:msd} shows the MSD for three different scenarios: (1)
$N_\alpha=100, N_\beta=0$, (2) $N_\alpha=0,N_\beta=100$, and (3)
$N_\alpha=50,N_\beta=50$. In all three the chemical gradient was set to be a
constant by setting the chemical diffusion, production and consumption to zero
($D_c = 0,k_\alpha=0,k_\beta=0$), and the initial profile equal to $c(\bfx) =
x$. The chemical concentration does not satisfy the periodic boundary conditions
in this case, however, its gradient is does and this is the only factor
influencing the simulation via the drift term on the $\beta$ particles. The
diffusion constant for each species was $D_\alpha=D_\beta=1$, and reactions
between particles types are turned off. For simulation (1) the $\alpha$
particles are not affected by the drift term and so the MSD is a straight line
with a low gradient. For simulation (2) the $\beta$ particles are affected by
the constant drift, which adds a dominant $t^2$ term to their MSD. For
simulation (3) the reactions are turned on again and there are both $\alpha$ and
$\beta$ cells, so the net effect on the MSD is a combination of these two
behaviours.

\section{Conclusions}

Particle-based models for biological processes have become of widespread use in mathematical biology. These come in many forms, with particle motions described by discrete or continuum random walks, complex interactions between particles (including reactions, hard-core interactions, or interaction potentials), and interactions with the environment (such as chemotaxis or transport through crowded or heterogeneous domains). In some cases, one also requires to couple a particle-based model, describing for example individuals cells, with a PDE model to represent a continuum field (such as a chemical concentration). In addition, despite their apparent simplicity, particle-based models can be challenging to implement and simulate, as they tend to scale badly with the number of particles in the system (which can be large in many applications) and, due to stochasticity, often many realisations of the same simulation are required. 

This diversity in particle-based models combined with the computational challenges in simulations, makes the implementation of particle-based models far from straightforward. In this paper we presented Aboria, a C\texttt{++} library, designed to provide the flexibility required to implement particle-based models commonly used in mathematical biology, in a high performance and easy to use fashion. 

We have demonstrated the usage of Aboria implementing a model for cell diffusion and chemotaxis with short-range interactions. The model has many of the features described above, namely, cells move according to biased Brownian motion, they interact with each other and with a chemical (that is modelled as a continuum), and there are reactions that change the number of particles. We have shown how Aboria can be used in combination with Python to produce outputs such as the cell densities, numbers, and the mean square displacement. 



\end{document}